\documentclass[oribibl, numbers]{llncs}

\usepackage{url}
\usepackage[colorlinks = true,
  linkcolor = blue,
  citecolor = blue,
  urlcolor = blue]{hyperref}

\usepackage{amsmath}
\usepackage{amssymb}
\usepackage{graphicx}
\usepackage{tabularx}
\usepackage{booktabs}
\usepackage{threeparttable}

\usepackage{fancyhdr} 
\fancypagestyle{firststyle}
{
\fancyhf{}
\fancyfoot[C]{\scriptsize{\textit{Applied Cybersecurity \& Internet Governance}, vol.~4, no.~1, pp.~1--18. This version is the authors' copy. The publisher's definite version is available online via \url{https://doi.org/10.60097/ACIG/213350}.}}
}

\begin{document}

\title{Vulnerability Coordination \\ Under the Cyber Resilience Act}

\author{Jukka Ruohonen\orcidID{\scriptsize{0000-0001-5147-3084}} \\
University of Southern Denmark
\vspace{15pt} \\
Paul Timmers\orcidID{\scriptsize{0000-0002-3446-583X}} \\
KU Leuven \& University of Oxford
\institute{}}

\maketitle

\begin{abstract}
  The Cyber Resilience Act (CRA) of the European Union (EU) imposes many new
  cyber security requirements practically to all network-enabled information
  technology products, whether hardware or software. The paper examines and
  elaborates the CRA's new requirements for vulnerability coordination,
  including vulnerability disclosure. Although these requirements are only a
  part of the CRA's obligations for vendors, also some new vulnerability
  coordination mandates are present. In particular, so-called actively exploited
  vulnerabilities require mandatory reporting. In addition to elaborating the
  reporting logic, the paper discusses the notion of actively exploited
  vulnerabilities in relation to the notion of known exploited vulnerabilities
  used in the United States. The CRA further alters the coordination practices
  on the side of public administrations. The paper addresses also these new
  practices. With the examination elaboration, and associated discussion based
  on conceptual analysis, the paper contributes to the study of cyber security
  regulations, providing also a few takeaways for further research.
\vspace{30pt} 
\end{abstract}

\section{Introduction}

\thispagestyle{firststyle} 

The EU's Cyber Resilience Act, that is, Regulation (EU) 2024/2847, was agreed
upon in October 2024. The paper presents an analysis of the CRA's implications
for vulnerability coordination, including vulnerability disclosure. Although the
handling of vulnerabilities is a classical research topic, the CRA imposes some
new requirements that warrant attention. It also enlarges the scope of
theoretical concepts involved. For these reasons, the paper carries both
practical relevance and timeliness. Regarding timeliness, the reporting
obligations upon severe incidents and actively exploited vulnerabilities will be
enforced already from September 2026 onward. Regarding practical relevance, it
is worth elaborating and clarifying the obligations and concepts involved
because some vendors have had difficulties with them due to a lack of expertise
and other reasons~\cite{Risto25}.  In terms of explicitly related work, the
paper continues the initial assessments \cite{Chiara22} that were conducted
based on the European Commission's proposal for the CRA. By using conceptual
analysis~\cite{Olsthroon17}, the research question examined is: what new
requirements the CRA entails for vulnerability coordination? To the best of the
authors' knowledge, the question has not been previously asked and answered.

The CRA is a part of a larger regulatory package for cyber security that was
pursued by the von der Leyen's first Commission, which was in office between
2019 and 2024. Although many of the other regulations and directives in the
package are on the side of critical infrastructure, the overall goal of improved
coordination, which can be understood broadly as a management of
dependencies~\cite{MaloneCrowston94, Ruohonen18IST}, is strongly present
throughout the package. Whether it is risk analysis or incident management, the
EU context generally implies that some coordination is required at multiple
levels; between cyber security professionals, companies, industry sectors,
national public authorities, and EU-level institutions, among other abstraction
levels~\cite{Ruohonen25COSE}. To make sense of the coordination involved, it is
useful to analytically separate horizontal coordination (such as between the
member states) and vertical coordination, which occurs between EU-level
institutions and units at a national level~\cite{Ruohonen22ICLR,
  Ruohonen24I3E}. Then, as will be elaborated, the CRA involves both horizontal
and vertical coordination. Coordination requires also technical infrastructures.

Regarding the larger regulatory package for cyber security, relevant to mention
is Directive (EU) 2022/2555, which is commonly known as the second network and
information systems (NIS2) security directive. With respect to vulnerability
coordination---a term soon elaborated in Section~\ref{sec: vulnerability
  coordination}, the directive strengthens and harmonizes the role of national
computer security incident response teams (CSIRTs) in Europe. In general,
various networks between CSIRTs, both in Europe and globally, have been
important for improving cyber security throughout the world, often overcoming
obstacles that more politically oriented cyber security actors have
had~\cite{Tanczer18}. Importantly for the present purposes, the NIS2 directive,
in its Article~12(2), specifies also an establishment of a new European
vulnerability database operated by the European Union Agency for Cybersecurity
(ENISA), which is the principal EU-level institution coordinating particularly
with the national CSIRTs but also involved in international
collaboration. Although the database is already online~\cite{ENISA25}, it is
still too early to say with confidence whether the CRA and NIS2 may affect the
infrastructures for vulnerability archiving, which is currently strongly built
upon institutions and companies in the United States, including particularly the
not-for-profit MITRE corporation involved in the assignment of Common
Vulnerabilities and Exposures (CVEs) and the National Vulnerability Database
(NVD) maintained by the National Institute of Standards and Technology (NIST) of
the United~States.

With respect to the EU's regulatory cyber security package, it can be further
noted that the CRA is rather unique in a sense that its legal background and
logic are motivated by the EU's product safety law and the general consumer
protection jurisprudence~\cite{Chiara22}. Therefore, also the associated
terminology is somewhat different compared to the other legislative acts in the
package. For instance, the CRA brings market surveillance authorities to the
public administration of cyber security in the EU. In general, these authorities
are responsible for ensuring that only compliant products enter into and
circulate on the EU's internal market; the examples include authorities ensuring
the safety of food, drugs, aviation, chemicals, and so forth.

The CRA covers the whole information technology sector and beyond. Both hardware
and software products with a networking functionality are in its scope. Among
the notable exemptions not covered by the regulation, as specified in the CRA's
Article~2, are medical devices, motor vehicles, and ships and marine
equipment. Furthermore, recital 12 clarifies that cloud computing is also
excluded. The CRA categorizes products into three categories: ``normal'',
``important'', and ``critical'' products. While all products need to comply with
the CRA's essential requirements, which have been analyzed in recent
research~\cite{Ruohonen25ESPREa, Ruohonen25ESPREb}, more obligations are placed
upon the important and critical products~ However, in what follows, the focus is
only on the CRA's implications for vulnerability coordination, including
vulnerability disclosure. With this focus in mind, the opening Section~\ref{sec:
  vulnerability coordination} introduces the vulnerability coordination and
disclosure concepts. Then, the subsequent Section~\ref{sec: cra requirements}
discusses and elaborates the CRA's new implications to vulnerability
coordination and disclosure. The implications are not uniform across everyone;
therefore, Section~\ref{sec: open source} continues by briefly elaborating the
requirements for open source software products and their projects. To some
extent, the CRA addresses also the pressing issues with supply chain
security. The important supply chain topic is discussed in Section~\ref{sec:
  supply chains}. A conclusion is presented in Section~\ref{sec:
  conclusion}. The final Section~\ref{sec: further work} discusses further
research.

\section{Vulnerability Disclosure and Coordination}\label{sec: vulnerability coordination}

There are various ways to disclose and coordinate a vulnerability between the
vulnerability's discoverer and a software vendor affected by the
vulnerability. Historically---and perhaps still so---a common practice was
so-called direct disclosure through which a discoverer and a vendor coordinate
and negotiate privately, possibly without notifying any further parties,
including maintainers of vulnerability databases and public
authorities~\cite{Ruohonen20CHB}. Partially due to the problems associated with
direct disclosure, another historical---and sometimes still practiced---process
was a so-called full disclosure via which a discoverer makes his or her
vulnerability discovery public, possibly even without notifying a vendor
affected or anyone else beforehand. One of the primary reasons for the full
disclosure practice was---and sometimes still is---a deterrence against vendors
who would not otherwise cooperate~\cite{Ruohonen20CHB, Syed25}. However, this
practice can be seen to increase security risks because sensitive information is
publicly available but no one has had time to react; there are no patches
available because a vendor may not even know about a vulnerability information
being available in the~public.

For this reason and other associated reasons, a third vulnerability disclosure
model emerged. It is called coordinated vulnerability disclosure; in essence, a
discoverer contacts an intermediary party who helps with the coordination and
collaboration with a vendor affected, often also providing advices to other
parties and the general public through security advisories and security
awareness campaigns. This coordinated disclosure model is the \textit{de~facto}
vulnerability disclosure practice today. Regarding the intermediaries involved,
a national CSIRT is typically acting as a coordinator, but also commercial
companies may act in a similar role, as is typical in the nowadays popular
so-called bug bounties and their online platforms. In addition to security
improvements, commercial bug bounty platforms may give vendors more control over
disclosure processes~\cite{Ahmed21}, while for discoverers they offer monetary
rewards, learning opportunities, and some legal safeguards~\cite{Akgul23}. It is
also important to emphasize that today practically all coordinated vulnerability
disclosure models further follow a so-called responsible disclosure, meaning
that vendors are given a fixed time period to develop and release patches before
information is released to the public.

With respect to Europe and the CSIRT-based coordinated vulnerability disclosure
model, the NIS2 directive specifies in its Article 12(1) three primary tasks for
the national CSIRTs in the member states: (1) they should identify and contact
all entities concerned, including the vendors affected in particular; (2) they
should assist natural or legal persons who report vulnerabilities; and (3) they
should negotiate disclosure timelines and arrange coordination for
vulnerabilities that affect multiple entities. While simple, these legal
mandates for public sector bodies are important because the legal landscape for
vulnerability disclosure has been highly fragmented in Europe~\cite{CEPS18,
  ENISA22}. The CRA relies partially upon these mandates in NIS2, but it also
imposes new requirements for both national CSIRTs and vendors.

Before continuing further, it should be clarified that a broader term of
vulnerability coordination has sometimes been preferred in the literature
because vulnerability disclosure presents only a narrow perspective on the
overall handling of vulnerabilities~\cite{Ruohonen18IST}. In addition to
contacting, communicating, assisting, and negotiating, numerous other tasks are
typically involved, including development of patches, testing these, and
delivering these to customers, clients, and users in general, assignment of CVE
identifiers, writing and releasing security advisories, archiving information to
vulnerability databases, and so forth and so on. These additional tasks do not
affect only vendors; also CSIRTs may need to carry out additional tasks beyond
those specified in NIS2. For these reasons, also the present work uses the
vulnerability coordination term.

\section{The CRA's Vulnerability Coordination Requirements}\label{sec: cra requirements}

The CRA separates voluntary vulnerability disclosure from mandatory obligations
placed upon vendors. Regardless whether a disclosure of a vulnerability is done
on voluntary or mandatory basis, it should be primarily done through a national
CSIRT, although voluntary disclosure may be done also to ENISA directly
according to the CRA's Article~15(1). Then, according to Article 15(2), both
vendors and others may also report not only vulnerabilities but also severe
security incidents and so-called near misses. These near misses are defined in
the NIS2's Article 6(5) as events that ``\textit{could have compromised the
  availability, authenticity, integrity or confidentiality of stored,
  transmitted or processed data or of the services offered by, or accessible
  via, network and information systems}'' but which were successfully
mitigated. These mitigated security events are one but not the only example of
new terminology brought by the EU's new cyber security law.

When a disclosure is done to a national CSIRT, the given CSIRT should then
without undue delay carry out vertical coordination toward the EU-level. This
important obligation placed upon national public sector authorities is specified
in the CRA's Article~16(1) according to which a common EU-level disclosure
infrastructure is established. It is maintained by ENISA. Through this
infrastructure, according to Article~16(2), a given national CSIRT should notify
other European CSIRTs designed as public sector authorities on those territories
on which a vendor's vulnerable products have been made available.

The territorial notion seems sensible enough when dealing with physical
products, whether routers, mobile phones, or automobiles. However, many digital
products, including software products in particular, are distributed online
without any specific territorial scope. For such products, it seems a reasonable
assumption that all European CSIRTs should be notified. This point applies
particularly to large software vendors. Similar reasoning was behind a schism
during the CRA's political negotiations because some companies and member states
feared that sensitive vulnerability information would unnecessarily spill over
throughout Europe due to the EU-level reporting
infrastructure~\cite{Ruohonen24I3E}. The schism was resolved by relaxing the
obligation placed upon CSIRTs.

In particular, according to Article~16(2), national CSIRTs may delay
coordination and reporting toward the EU-level under exceptional circumstances
requested by a vendor affected. The article specifies three clarifying
conditions regarding these exceptional circumstances: (1) a given vulnerability
is actively exploited by a malicious actor and no other member state is
affected; (2) immediate reporting would affect the essential interests of a
member state; and (3) imminent high cyber security risk is stemming from
reporting. Although a national CSIRT should always still notify ENISA about its
decision to delay reporting, it seems that the three conditions are so loose
that most cases could be delayed in principle. Time will tell how the regulators
interpret the exceptional circumstances in practice.

When no exceptional circumstances are present and reporting is done normally by
a CSIRT, it should be clarified that both vertical and horizontal coordination
is present. A CSIRT's notification toward the EU-level is vertical in its
nature, while distributing information to other CSIRTs is a horizontal
activity. In addition, further horizontal coordination is present because the
CRA's Article~16(3) specifies that a CSIRT should also notify any or all market
surveillance authorities involved. This obligation complicates the coordination
substantially because it may be that numerous market surveillance authorities
are present, including everything from product safety authorities to
customs. Alternatively, it may be that a member state has specified its national
CSIRT also as a market surveillance authority in the CRA's context. Again, time
will tell how the coordination works in practice. Future evaluation work is
required later in this~regard.

The exceptional circumstances noted contain the concept of actively exploited
vulnerabilities. Such vulnerabilities are a core part of the CRA. An actively
exploited vulnerability is defined in Article~3(42) as ``\textit{a vulnerability
  for which there is reliable evidence that a malicious actor has exploited it
  in a system without permission of the system owner}''. Recital 68 clarifies
that security breaches, including data breaches affecting natural persons, are a
typical example. The recital also notes that the concept excludes good faith
testing, investigation, correction, or disclosure insofar as no malicious
motives have been involved.

Then, according to Article~14(1), all vendors should notify about actively
exploited vulnerabilities in their products to both a given national CSIRT and
ENISA. Article~14(2) continues by specifying that vendors should deliver an
early notification within 24 hours, an update within 72 hours, and a final
report within two weeks. As could be expected, concerns have been expressed
among vendors about these strict deadlines~\cite{Risto25}. Also reporting
quality remains a question mark~\cite{Ruohonen25COSE}. In any case, the
subsequent paragraphs 3 and 4 in Article~14 obligate vendors to report not only
about actively exploited vulnerabilities but also about severe security
incidents. These are defined in the NIS2's Article~6(6) analogously to the near
misses but without the successful mitigation condition. Furthermore,
Article~14(8) specifies that both actively exploited vulnerabilities and severe
security incidents should be communicated to the users impacted, possibly
including the general public, either by a vendor affected or a given~CSIRT.

\begin{figure}[th!b]
\centering
\includegraphics[width=\linewidth, height=7cm]{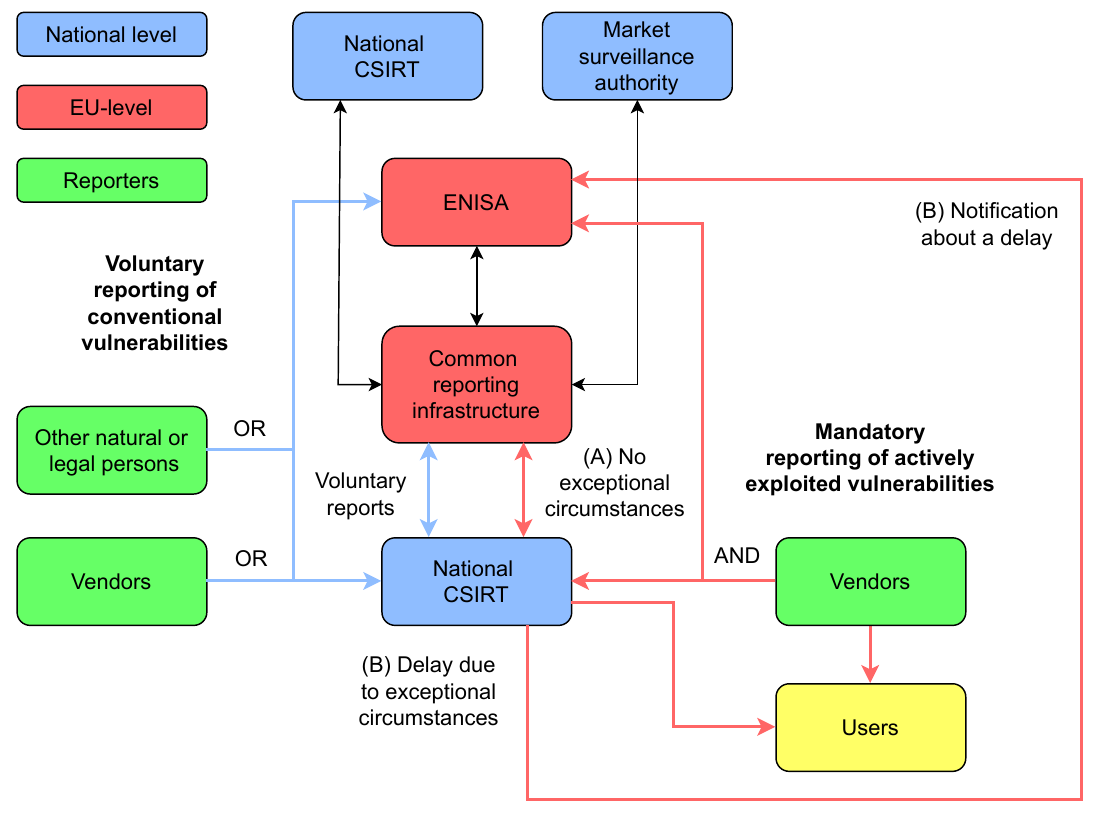}
\caption{Vulnerability Coordination Under the CRA in an Analytical Nutshell}
\label{fig: reporting}
\end{figure}

Regarding vulnerabilities, as can be seen also from Fig.~\ref{fig: reporting},
it should be underlined that the mandatory notifications only involve actively
exploited vulnerabilities. Thus, among other things, so-called silent patching
is excluded from the CRA's scope; it refers to cases in which a vendor has found
a vulnerability from its product and silently corrected it without making the
information available to others~\cite{Ruohonen20CHB}. A more important point is
that the definition for actively exploited vulnerabilities is rather loose. When
considering a possibility of underreporting---a topic that has been discussed in
the literature in somewhat similar settings~\cite{Gorton16}, particularly the
wording about reliable evidence raises a concern. A relevant question is: whose
evidence? If vendors alone answer to the question, underreporting may
occur. Though, other parties too, including public authorities, have threat
intelligence platforms, security information and event management (SIEM)
systems, and many related techniques deployed. Against this backdrop, a vendor
deciding not to report an actively exploited vulnerability may later on find
itself as having been non-compliant. In addition, Article~60 allows market
surveillance authorities to conduct sweeps to check compliance.

It is important to further note that the concept of actively exploited
vulnerabilities is used implicitly also in the United States. In particular, the
Cybersecurity \& Infrastructure Agency (CISA) therein maintains a catalog of
known exploited vulnerabilities (KEVs)~\cite{CISA24}. Given the ENISA's and
CISA's established cooperation~\cite{ENISA23}, it~remains to be seen whether the
catalog will be used also for enforcing the CRA or otherwise helping its
implementation. An analogous point applies to the international coordination
between vulnerability databases, including particularly between the NVD and the
EU's new vulnerability database. A related important point is that less strict
vulnerability reporting requirements are imposed upon open source software
projects and their supporting foundations, which are known as stewards in the
CRA's parlance.

Regarding KEVs, the available documentation~\cite{CISA24} says that three
conditions must be satisfied for a vulnerability to be a KEV. First, a KEV must
have a CVE identifier assigned. Then, exploitability is understood along the
lines of the Common Vulnerability Scoring System (CVSS) as the ease of an
attacker to take advantage of a CVE-referenced vulnerability. Thus,
exploitability varies according to a number of factors, such as whether a
vulnerability can be exploited remotely through a network, whether a
proof-of-concept exploit is available, whether authentication or privileges are
required for exploitation, and so forth and so on. A vulnerability that cannot
be exploited under reasonable constraints cannot be a KEV. Second, however, for
an exploitable vulnerability to be a KEV, it must have been either exploited or
under active exploitation Both terms require that there is ``reliable evidence
that execution of malicious code'' has occurred~\cite{CISA24}. Such evidence
does not necessarily mean that a vulnerability has been successfully exploited;
a KEV can refer also to an attempted exploitation, including those conducted in
a honeypot or related deception environment. It can be also noted that execution
of malicious code rules out network scanning and denial of service
attacks. Third, a KEV must have clear remediation guidelines for it to be
accepted into the catalog. These range from updates provided by vendors to
temporary mitigations and workarounds. It will be interesting to see whether
also the CRA's concept of actively exploited vulnerabilities will be tied to
these or closely related principles or whether a new European conceptual
framework will be developed in the future.

Finally, it is important to emphasize that the CRA's so-called essential cyber
security requirements, which, like with the EU's product safety
law~\cite{Ruohonen22ICLR}, will be accompanied with harmonized standards,
contain also obligations explicitly related to vulnerabilities and their
coordination. In general, as specified in Annex~I, vendors should only make
available products without known exploitable vulnerabilities, but they should
also ensure that vulnerabilities can be addressed through security updates,
including, where applicable, automated security updates separated from
functional updates that bring new features~\cite{Ruohonen25ESPREb}. They must
also establish and enforce a coordinated vulnerability disclosure policy, and
apply effective and regular security testing and reviews, among other things. As
soon discussed in Section~\ref{sec: supply chains}, the essential cyber security
requirements imply also new requirements for supply chain management.

\section{The CRA's Vulnerability Coordination Requirements for Open Source Projects}\label{sec: open source}

Many open source software (OSS) projects and their foundations were critical
about the CRA during its policy-making. Some notable OSS foundations even raised
an argument that online distribution of OSS might be blocked in the EU due to
the CRA~(see~\cite{LF23}, among many others during the CRA's
policy-making). While the argument might have merely been a part of an
aggressive lobbying strategy, some have used a term ``regulatory anger'' to
describe the situation during the political negotiations, concluding that the
new regulatory obligations for OSS projects are modest at best and relatively
easy to fulfill~\cite{Hubert23}. The reason for the only modest implications
originates from the fact that the CRA essentially only applies to commercial OSS
projects. This restriction is clarified in recitals~17, 18, and 19. These
recitals clarify that natural or legal persons who merely contribute to OSS
projects are not within the regulation's scope. Nor is a mere distribution of
OSS in the CRA's scope. Instead, the keyword is monetization.

To quote from recital 19, the CRA only applies to OSS projects that
``\textit{are ultimately intended for commercial activities}'', including
``\textit{cases where manufacturers that integrate a component into their own
  products with digital elements either contribute to the development of that
  component in a regular manner or provide regular financial assistance to
  ensure the continuity of a software product}''. This clarification essentially
means that the main target is OSS foundations; the Linux Foundation would be the
prime example. However, none of the keywords---including monetization, ultimate
intentions, regular contributions, and financial assistance---are actually
defined in Article~3. This lack of definitions makes it unclear how the
regulators and legal systems will interpret the situation. Regardless, even with
vague definitions and potential disputes, the actual regulatory mandates placed
upon OSS foundations are only light.

Thus, Article 24 specifies three obligations for OSS foundations: (1) they are
obliged to develop and publish cyber security policies that address particularly
the coordination of vulnerabilities and foster the voluntary reporting of
vulnerabilities; (2) they must cooperate with market surveillance authorities
who may also request the cyber security policies; (3) and they too must report
about actively exploited vulnerabilities and severe security incidents. While
the first two obligations are mostly about documentation, and most large OSS
foundations likely already have such documented policies in place, the reporting
obligation may perhaps be somewhat difficult to fulfill because the supporting
OSS foundations are mostly involved in the development---and not deployment---of
open source software. This point reiterates the earlier remark about reliable
evidence. However, according to Article~64(10)(b), OSS foundations are excluded
from the scope of administrative fines. This exclusion casts a small doubt over
whether the supervising authorities have a sufficient deterrence against
non-compliance.

It can be mentioned that the initial legislative proposal for the CRA,
identified as COM/2022/454 final, was more stringent about OSS. Thus, perhaps
the lobbying by OSS stakeholders was also successful. Having said that, there
are two important points that many OSS lobbyists missed during the
negotiations. The first is that the CRA's Article~25 entails a development of
voluntary security attestation programmes for OSS projects to ensure compliance
with the essential cyber security requirements. Such programmes can be
interpreted to boost the security of open source software~\cite{Hubert23}. The
second point stems from Article 13(6), which mandates that a vendor who
discovers a vulnerability from a third-party software component, including an
OSS component, must report the vulnerability to the natural or legal persons
responsible for the component's development or maintenance. Also this reporting
obligation can be seen to improve the security of open source software too. It
is also directly related to supply~chains.

\section{The CRA and Supply Chains}\label{sec: supply chains}

Both the CRA and NIS2 directive address supply chain security, either explicitly
or implicitly. In fact, throughout the CRA's motivating 130 recitals, supply chain
security is frequently mentioned. Many of the new supply chain obligations are
specified in Annex I. Accordingly, vendors should try to identify and document
vulnerabilities in third-party software components they use; for this task, they
are mandated to draw up a software bill of materials (SBOM) for their
products.

A SBOM is defined in Article~3(39) to mean ``\textit{a formal record containing
  details and supply chain relationships of components included in the software
  elements of a product with digital elements}''. In practice, SBOMs are
primarily about software dependencies, whether managed through a package manager
or bundled into a software product directly. These are also under active
research~\cite{Bi24, ODonoghue24}. As motivated by the CRA's recital~77, SBOMs
are seen not only important for regulatory purposes but they are also perceived
as relevant for those who purchase and operate software products. However, the
same recital notes that vendors are not obliged to make SBOMs public, although
they must supply these to market surveillance authorities upon request according
to Article~13(25). Confidentiality must be honored in such~cases.

Yet, SBOMs are not the only new obligation related to supply chains.
Historically, there was a problem for many vulnerability discoverers to find
appropriate contact details about organizations and their natural persons
responsible~\cite{Ruohonen20CHB}. The problem and related obstacles likely still
persist today, as hinted by experiences with large-scale coordinated
vulnerability disclosure and the poor adoption rate of new initiatives such as
the so-called~\texttt{security.txt} for the world wide web~\cite{Chen24,
  Hilbig24, Stock16}. To this end, paragraph 6 in Annex I mandates vendors to
also provide contact details for reporting vulnerabilities. This obligation is
important also for regulators; as was noted in Section~\ref{sec: vulnerability
  coordination}, they should coordinate vulnerability handling, including
disclosure, also in case multiple parties are involved. Without knowing whom to
contact, let alone which components have been integrated, such multi-party
coordination is practically impossible. The same point applies to the earlier
note about obligations placed upon vendors who may discover vulnerabilities in
third-party components they use. The obligation to provide contact details is
recapitulated in paragraph 2 of Annex~II.

\section{Conclusion}\label{sec: conclusion}

The paper examined, elaborated, and discussed the implications from the EU's new
CRA regulation upon vulnerability coordination. When keeping in mind the paper's
framing only to vulnerability coordination, the CRA's new obligations cannot be
argued to be substantial---in fact, it can be argued that many of these
obligations have already long belonged to security toolboxes of responsible
vendors prioritizing cyber and software security. Nevertheless, (1)~among the
new legal obligations are mandatory reporting of actively exploited
vulnerabilities and severe cyber security incidents for which (2) strict
reporting deadlines are also imposed. On the public administration side, (3)~the
CRA strengthens the EU-level administration due to ENISA's involvement and the
establishment of a common European reporting infrastructure, including a new
European vulnerability database. In addition, (4)~administrative complexity is
also increased due to the involvement of new public sector bodies, including
market surveillance authorities in particular.  Furthermore, (5) the new
mandatory reporting obligations apply also to OSS foundations with commercial
ties, but also commercial vendors must report vulnerabilities to OSS projects
whose software they use. Finally, (6) the CRA imposes new requirements also for
supply chain management, some of which are related to vulnerability
coordination; the notable examples are mandates to roll out SBOMs and provide
contact details.

\section{Further Work}\label{sec: further work}

Of the new regulatory obligations and points raised thereto, further research is
required about the concept of actively exploited vulnerabilities. Such research
might also generally reveal whether the existing empirical vulnerability
research has been biased in a sense that serious vulnerabilities affecting
actual deployments have perhaps been downplayed; with some notable
exceptions~\cite{Bilge12}, the focus has often been on reported vulnerabilities
affecting software products, components, and vendors, not on their actual
deployments~\cite{Paschenko18}. Another related topic worth examining is the
theoretical and practical relation between KEVs and the CRA's notion of actively
exploited vulnerabilities.

On a more theoretical side of research, further contributions are required also
regarding the other fundamental concepts involved. Regarding the exploitation
theme, many further concepts have been introduced in recent research, among them
a concept of likely exploited vulnerabilities~\cite{Mell25}. A conceptual
enlargement applies also to security incidents. Although attempts have been made
to better understand and theorize how different incident types are related to
each other in terms of EU law~\cite{Ruohonen25COSE}, it still remains generally
unclear what exactly separates a \textit{severe} incident from a
``conventional'', non-severe incident or a \textit{significant} incident from a
severe incident. These and other related interpretation difficulties can be seen
also as a limitation of the current paper.

Then, regarding supply chains, further research is required also on
SBOMs. Although technical research is well underway in this regard, including
with respect to a question of how to generate SBOMs
automatically~\cite{Rabbi24}, many questions remain about applicability,
usability, and related aspects, including for regulatory purposes. Although
perhaps only partially related to the paper's explicit theme, a relevant
research question would also be whether, how, and how well vendors can comply
with the requirement to provide security patches, preferably in an automated
manner. When recalling the CRA's scope, the question is non-trivial. While
providing automated security updates for Internet of things devices sounds like
a reasonable and well-justified goal, the situation may be quite different in
industrial settings~\cite{Ruohonen25ESPREb}. It also remains generally unclear
whether the CRA applies to physical security, including patching of software
operating in air-gapped and related deployments. As always, further
policy-oriented research is also needed. Among other things, a close eye is
needed to monitor and evaluate the regulation's future implementation, adoption,
administration, and enforcement.

Regarding the CRA's future implementation, a particularly relevant research
topic involves examining the upcoming harmonized standards upon which the
essential cyber security requirements are based. This topic is important already
because complying with these standards provides a presumption of conformity
according to the CRA's Article~27(1). The topic can be extended to other
conformity assessment procedures involved, including the mandatory third-party
audits required for critical products and the CRA's relation to the older cyber
security certification schemes in the EU. Among other things, a good topic to
examine would be the accreditation of conformity assessment bodies. It may be
that the CRA will also entail a new labor market demand for cyber security
professionals specialized into security testing, auditing, and related
domains. At the same time, as also implicitly remarked in the CRA's Article~10,
there is allegedly a cyber security skill gap in Europe and
elsewhere~\cite{Furnell17}. As the gap may correlate with
education~\cite{Ricci24, Zivanovic24}, it may also be worthwhile to further
study the CRA and its future implementation with education and curricula in
mind. Regarding adoption, it is relevant to know which conformity assessment
procedures vendors will prefer in the future; according to Article~32(1), there
are four distinct procedures for conformity assessments under the CRA.

Although it is too early to speculate what the CRA's broader implications may
be, three general points can be still raised. First, the larger regulatory cyber
security package in the EU has expectedly raised some criticism that
fragmentation and complexity have increased~\cite{deVasconcelosCasimiro23,
  Ruohonen24I3E}. The CRA has contributed to this increase. The CRA also
interacts with the EU's other legal cyber security acts. In particular, as the
NIS2 directive also imposes reporting obligations for vendors, synchronization
between the CRA and NIS2 has been perceived as crucial~\cite{Chiara22}. Second,
a notable limitation of the CRA is that it offers no legal guards against
criminal or civil liability from disclosing vulnerabilities. Such guards have
often been seen important in the literature~\cite{Akgul23, CEPS18}. Third, it
remains to be seen whether the CRA actually improves the cyber security of
products. Although all points require further research, including rigorous
evaluations, particularly the last point is fundamental in terms of the CRA's
explicit goals. According to the Act's first recital, cyber insecurity affects
not only the EU's economy but also its democracy as well as the safety and
health of consumers. Already against this backdrop, it is important and relevant
to evaluate in the future how well the CRA managed to remove or at least
mitigate cyber insecurity in the EU.

\enlargethispage{1.2cm}
\bibliographystyle{splncs03}

\end{document}